\documentclass[aps,preprint]{revtex4}

\begin{document}

\bibliographystyle{prsty}

\draft

%\wideabs{

\title{ Quantitative Measure of Stability in Gene Regulatory Networks }
\author{ P. Ao }
\address{ Department of Mechanical Engineering,
         University of Washington, Seattle, WA 98195, USA    }
%\address{ }
\date{ July 6, 2005 }
%\date{\today}

%\maketitle

\begin{abstract}
A quantitative measure of stability in stochastic dynamics starts
to emerge in recent experiments on bioswitches. This quantity,
similar to the potential function in mathematics, is deeply rooted
in biology, dated back at the beginning of quantitative
description of biological processes: the adaptive landscape of
Wright (1932) and the development landscape of Waddington (1940).
Nevertheless, its quantitative implication has been frequently
challenged by biologists. Recent progresses in quantitative
biology begin to meet those outstanding challenges.
\end{abstract}

%\pacs{PACS numbers: }

%}

\maketitle

With the successful experimental work and theoretical analysis on
simple artificial gene networks [1],  Acar et al  [2] further
explored the quantitative behaviors in a living gene regulatory
network, the yeast galactose-signaling network. Their work
provides a fine and new example of quantitative understanding of
the stability and reversibility of cellular differentiation state
in terms of the "potential" or "energy" landscape.  I wish to
point out here that together with the extensive, and perhaps more
quantitative, study in another living genetic regulatory network
not discussed by Acar et al, the phage lambda [3,4,5], where
similar conclusion was obtained [6], a powerful theoretical
modelling framework of stochastic dynamics in biological networks
begins to emerge.

There has been a long tradition in biology to understand the
stability problem in terms of "potential" landscape also not
mentioned by Acar et al. In 1932 S. Wright proposed the adaptive
landscape in evolution in the context of genetics [7]. In 1940
C.H. Waddington proposed the developmental landscape to understand
the stability and differentiation in developmental biology [8]. In
the context directly linked to bio-switching phenomena M. Delbruck
proposed in 1949 that the potential landscape could be used for
the modelling [9].  Unfortunately, despite such a continuous
effort in biology and the fact that such a landscape concept have
been permeated into other fields such as physics and engineering,
the potential landscape has been at best viewed in biology as a
useful metaphor. It has been generally believed that this concept
could not be quantified [10]. It should be instructive to mention
that in the general setting of nonequilibrium processes,
stochastic dynamical processes in life sciences as discussed in
[2] and [6] are obviously belong to this category, the search for
such a quantitative potential landscape has been performed during
past a few decades and has only been partially successful [11].
Such a difficulty may also be reflected in the debate in
mathematics around the catastrophe theory, such as vector vs
gradient fields [12].

Those recent studies [1,2,6] reveal that a potential landscape
concept is not only metaphoric, also quantitative. An attempt to
put in this concept into a rigorous mathematical framework has
been under way [13], though various obstacles still lie ahead. It
appears that the time is ripe to consider the theoretical
modelling from another perspective deeply rooted in biology.

{\ }

\noindent{\bf Acknowledgement:}
This work was supported in part by a USA NIH grant under HG002894.

{\ }

{\bf References} \\
1. M. Kaern, T.C. Elston, W.J. Blake, and J.J. Collins (2005)
Stochasticity in gene  expression: from theories to phenotypes.
Nature Reviews: Genetics 6:  451-464.  \\
2. M. Acar, A. Becskei, and A. van Oudenaaden (2005) Enhancement of cellular memory
    by reducing stochastic transitions.  Nature 435: 228-232. \\
3. J.W. Little, D.P. Shepley, and D.W. Wert (1999) Robustness of a
gene regulatory circuit. EMBO J. 18:4299-4307.  \\
4. M. Ptashne (2004) A Genetic Switch: phage lambda revisited. 3rd
edition. Cold Spring
      Harbor Laboratory Press, Cold Spring Harbor.  \\
5. I.B. Dodd, K.E. Shearwin and J.B. Egan (2005) Revisited gene
regulation in
       bacteriophage $\lambda$. Curr. Opin. Gen. Devlop. 15: 145-152.  \\
6. X.-M. Zhu, L. Yin, L. Hood, and P. Ao (2004) Robustness,
stability and efficiency of
     phage lambda genetic switch: dynamical structure analysis.
     J. Bioinform. Comput. Bio. 2: 785-817.
7. S. Wright (1932) The roles of mutation, inbreeding,
crossbreeding and selection in
     evolution. Proceedings of the 6th International Congress of Genetics. 1:
     356-366. \\
8. C.H. Waddington (1940) Organizers and Genes. Cambridge
University Press,  Cambridge. \\
9. M. Delbruck (1949) C.N.R.S., Paris, pp33. The English
translation by C.H. Waddington was in Ref.[12], pp29-31.  \\
10. P. Ao (2005) Laws in Darwinian evolutionary theory.
    Physics of Life Reviews 2: 117-156.  \\
11. G. Nicolis and I. Prigogine (1977) Self-Organization in
Nonequilibrium Systems: from
     dissipative structure to order through fluctuations. Wiley, New
     York. \\
12. R. Thom (1983) Mathematical Models of Morphogenesis.
      Wiley, New York.  \\
13. C. Kwon, P. Ao, D.J. Thouless (2005) Structure of stochastic
dynamics near fixed
      point. Submitted to PNAS (cond-mat/0506280).

\end{document}